\documentclass[showpacs]{revtex4} 
\usepackage{amsmath,amssymb,graphicx,textcomp,upgreek} 
\usepackage{natbib}
\usepackage[english]{babel}

\begin{document}
\vspace{0mm}
\title{NUCLEAR AND ELECTRONIC COHERENCE IN SUPERFLUID HELIUM} %
\author{Yu.M. Poluektov}
\email{yuripoluektov@kipt.kharkov.ua} %
\affiliation{National Science Center ``Kharkov Institute of Physics
and Technology'', 1, Akademicheskaya St., 61108
Kharkov, Ukraine} %

\begin{abstract}
A semi-phenomenological model of a many-particle system of $^4$He
atoms is proposed, in which a helium atom is considered as a complex
consisting of a nucleus and a bound pair of electrons in the singlet
state. At zero temperature, there are two Bose-Einstein condensates
of particles with opposite charges, namely, a condensate of
positively charged nuclei and a condensate of negatively charged
electron pairs. It is shown that in such a system there exist two
excitation branches: sound and optical. On the basis of this model
an interpretation of experiments on the study of the electrical
activity of superfluid helium is proposed. The frequency at which
the resonant absorption of a microwave radiation is observed is
interpreted as a gap in the optical branch. It is shown that the
distribution of the electric potential in a standing wave in a
resonator is similar to that observed experimentally.
\newline%
\end{abstract}
\maketitle

\section{Introduction}\vspace{-0mm} 
In the experiments of Rybalko [1--7] with superfluid helium, there
were registered effects which demonstrate an increased electrical
activity of this neutral medium. The activity manifests itself both
at low frequencies in sound and torsion experiments [1,2] and at
high frequencies in the interaction with a microwave radiation
[3--7]. In one group of effects, the electrical oscillations were
observed under fluctuations of temperature $T$  [1] and under
oscillations of the difference of the superfluid and normal
velocities ${\bf w}={\bf v}_s-{\bf v}_n$ [2]. In experiments of
another type, the resonant absorption of a microwave radiation was
found [3--7] at a frequency close to 180 GHz. These results were
mainly confirmed in later experiments [8--12].

Until now there have been carried out a significant number of
theoretical works where attempts have been made to explain the
observed effects. However, it seems unlikely that such effects can
be explained while remaining within the framework of the traditional
theory of superfluidity, where the internal structure of atoms is
not taken into account. The internal structure was taken into
account in theoretical works [13,14], in which particles were
considered as hydrogen-like atoms. In this work, we propose a
semi-phenomenological model of a superfluid system of particles
whose structure is closer to the real structure of the helium atom.

Before experiments in which the electrical activity was discovered,
in the theoretical study of the superfluid properties of liquid
helium atoms were usually considered as structureless particles with
zero spin. At zero temperature, a system of $N$ atoms obeying the
Bose statistics is described by the wave function $\Psi({\bf
r}_1,{\bf r}_2,\ldots,{\bf r}_N)$ being symmetric with respect to
the permutations of position vectors ${\bf r}_j$. In a low-density
system, when all particles are in the same state, the total wave
function can be represented as a product of identical functions
$\psi({\bf r}_1)\psi({\bf r}_2)\ldots\psi({\bf r}_N)$ characterizing
the state of an individual particle in the condensate. The function
$\psi({\bf r})$ obeys the well-known Gross-Pitaevskii equation
[15,16]. Such state is coherent [13]. For dense systems the
structure of the symmetric wave function proves to be more complex,
and in this case an important role is also played by pair
correlations and correlations of a larger number of particles
[17,18]. In this work we will not touch upon the question of the
role of higher correlations. Thus, in contrast to the model of an
ideal Bose gas where the condensate particles actually fall out of
consideration, when taking into account the interparticle
interaction the condensate particles are described by some effective
complex wave function $\psi({\bf r})$. We will call the condensate
of interacting particles as the coherent Bose-Einstein condensate.
The concept of the superfluid component of He\,II as a superposition of
oppositely charged coherent boson condensates -- nuclear and
electron -- was considered in work [19], where the cause of generation of
an electric field was associated with the acceleration
of electrons and nuclei which have very different masses.

A helium atom consists of a nucleus (alpha particle) with zero spin
and a pair of electrons. In the ground state the spins of electrons
are directed oppositely, so that the total spin of a pair is zero
(parahelium). A pair of electrons in such a singlet state is a very
strong formation. In order to transfer a pair of electrons from the
singlet state to the triplet state with the total spin equal to
unity (orthohelium), an energy of 19.8\,eV should be spent,
and the energy of the first excited state of parahelium is 20.6\,eV
higher than that of the ground state. This
makes it possible to consider a pair of electrons in the ground
state of the helium atom as a single object resembling a Cooper pair
localized near a nucleus. On this basis, in the proposed model the
helium atom will be considered as a complex consisting of a spinless
nucleus with charge $2|e|$ and a particle with zero spin and charge
$-2|e|$.

When taking into account the internal structure of the atom, both
nuclei and pairs of bound electrons pass into the condensate. Thus,
this model considers a neutral system of two Bose-Einstein
condensates of nuclei and electron pairs with opposite charges. The
fluctuations of the densities of the number of particles in
condensates are accompanied by the fluctuations of the densities of
charge, current and electric potential. This article studies small
oscillations of such a system of two condensates and shows that
there exist two branches of elementary excitations -- the sound
branch and the optical branch. It is also shown that the
distribution of the electric potential in a standing wave in a
resonator coincides with the distribution observed in the experiment
[20].

Based on the analysis of the proposed model, it was concluded that
the electrical effects observed in superfluid helium are a
consequence of the perturbation of its coherent system determining
the value of the superfluid density. There are three parameters that
lead to a change in the superfluid density: temperature, superfluid
flow and pressure. Estimates show that the largest perturbation of
the coherent system is induced by the temperature fluctuations. A
somewhat smaller effect is caused by the fluctuations of the
superfluid flow. The least influence on the coherent system is
exerted by the pressure fluctuations.\vspace{-1mm}

\section{Dynamical equations of the coherent system of nuclei  and electron pairs}\vspace{-2mm}
In the secondary quantization representation, the system of nuclei
will be described by the field operator $\psi_\alpha({\bf r},t)$ and
the system of pairs of bound electrons by the field operator
$\psi_e({\bf r},t)$. These operators obey the usual commutation
relations
\begin{equation} \label{01}
\begin{array}{l}
\displaystyle{%
  \big[\psi_\alpha({\bf r},t),\psi_\alpha^+({\bf r}',t)\big]=\delta({\bf r}-{\bf r}'), \quad     %
  \big[\psi_\alpha({\bf r},t),\psi_\alpha({\bf r}',t)\big]=0,      %
}\vspace{2mm}\\ %
\displaystyle{%
  \big[\psi_e({\bf r},t),\psi_e^+({\bf r}',t)\big]=\delta({\bf r}-{\bf r}'), \quad     %
  \big[\psi_e({\bf r},t),\psi_e({\bf r}',t)\big]=0      %
}%
\end{array}
\end{equation}
and commute with each other. The Hamiltonian has the form $H=H_K+H_I+H_E$, where %
\begin{equation} \label{02}
\begin{array}{l}
\displaystyle{%
  H_K=-\int\!d{\bf r}\left\{\psi_\alpha^+({\bf r})\!\left[\frac{\hbar^2}{2M}\Delta+\mu_\alpha\right]\!\psi_\alpha({\bf r})+\psi_e^+({\bf r})\!\left[\frac{\hbar^2}{2m}\Delta+\mu_e\right]\!\psi_e({\bf r})\right\}, %
}
\end{array}
\end{equation}
\vspace{-3mm}%
\begin{equation} \label{03}
\begin{array}{l}
\displaystyle{%
  H_I=\frac{1}{2}\int\!d{\bf r}d{\bf r}'\Big\{\psi_\alpha^+({\bf r})\psi_\alpha^+({\bf r}')U_{\alpha\alpha}(|{\bf r}-{\bf r}'|)\psi_\alpha({\bf r}')\psi_\alpha({\bf r})\,+ %
}\vspace{2mm}\\ %
\displaystyle{%
   \hspace{24mm}+\,\psi_e^+({\bf r})\psi_e^+({\bf r}')U_{ee}(|{\bf r}-{\bf r}'|)\psi_e({\bf r}')\psi_e({\bf r})\,+%
}\vspace{2mm}\\ %
\displaystyle{%
   \hspace{24mm}+\,2\psi_\alpha^+({\bf r})\psi_\alpha^+({\bf r})U_{\alpha e}(|{\bf r}-{\bf r}'|)\psi_e({\bf r}')\psi_e({\bf r}')\Big\}, %
}%
\end{array}
\end{equation}
\vspace{-4mm}%
\begin{equation} \label{04}
\begin{array}{l}
\displaystyle{%
  H_E=\int\!d{\bf r}\left\{|e|\!\left[\psi_\alpha^+({\bf r})\psi_\alpha({\bf r})-\psi_e^+({\bf r})\psi_e({\bf r})\right]\!\varphi({\bf r})+\frac{\big(\nabla\varphi({\bf r})\big)^2}{8\pi}\right\}. %
}
\end{array}
\end{equation}
Here $M,m$ are the effective masses of a nucleus and an electron
pair, $e$ is the electron charge. Note that the effective masses in
a many-particle system of interacting particles do not have to
coincide with the mass of a helium nucleus $M_\alpha$ and the mass
of a pair of free electrons $2m_e$, but they are phenomenological
parameters. For definiteness we will assume that $M>m$. The electric
field is taken into account in the nonrelativistic approximation
through the scalar potential $\varphi({\bf r})$. For simplicity, in
the following we choose the interaction potentials in the delta-like form: %
\begin{equation} \nonumber
\begin{array}{l}
\displaystyle{%
  U_{\alpha\alpha}(|{\bf r}-{\bf r}'|)\equiv g_\alpha\delta({\bf r}-{\bf r}'), \quad  %
  U_{ee}(|{\bf r}-{\bf r}'|)\equiv g_e\delta({\bf r}-{\bf r}'), \quad  %
  U_{\alpha e}(|{\bf r}-{\bf r}'|)\equiv g_{\alpha e}\delta({\bf r}-{\bf r}').  %
}
\end{array}
\end{equation}
We assume that $g_\alpha>0$, $g_e>0$, $g_{\alpha e}<0$. The
operators of the number of nuclei and the number of electron pairs,
respectively, are
\begin{equation} \label{05}
\begin{array}{l}
\displaystyle{%
  N_\alpha=\int\!d{\bf r}\,\psi_\alpha^+({\bf r})\psi_\alpha({\bf r}),\quad N_e=\int\!d{\bf r}\,\psi_e^+({\bf r})\psi_e({\bf r}).  %
}
\end{array}
\end{equation}
In the Heisenberg representation, the operators depend on time and
obey the equations of motion
\begin{equation} \label{06}
\begin{array}{l}
\displaystyle{%
  i\hbar\frac{\partial \psi_\alpha({\bf r},t)}{\partial t}=\big[\psi_\alpha({\bf r},t),H\big], \quad  %
  i\hbar\frac{\partial \psi_e({\bf r},t)}{\partial t}=\big[\psi_e({\bf r},t),H\big].  %
}
\end{array}
\end{equation}
Using the formulas (1)\,--\,(4), we obtain an explicit form of
equations for the field operators. In accordance with the fact that
at temperatures close to zero most Bose particles are in a single
state, by analogy to the Gross-Pitaevskii approach [15,16] one can
neglect the commutation properties of the operators and consider
them as ordinary functions. As a result, we obtain the equations
\begin{equation} \label{07}
\begin{array}{l}
\displaystyle{%
  i\hbar\frac{\partial \psi_\alpha}{\partial t}=-\!\left(\frac{\hbar^2}{2M}\Delta+\mu_\alpha+|e|\varphi\right)\psi_\alpha+ g_\alpha|\psi_\alpha|^2\psi_\alpha+g_{\alpha e}|\psi_e|^2\psi_\alpha,  %
}%
\end{array}
\end{equation}
\vspace{-4mm}%
\begin{equation} \label{08}
\begin{array}{l}
\displaystyle{%
  i\hbar\frac{\partial \psi_e}{\partial t}=-\!\left(\frac{\hbar^2}{2m}\Delta+\mu_e-|e|\varphi\right)\psi_e+ g_e|\psi_e|^2\psi_e+g_{\alpha e}|\psi_\alpha|^2\psi_e.  %
}%
\end{array}
\end{equation}
Equating to zero the variation of the energy with respect to the
scalar potential, we arrive at the Poisson equation
\begin{equation} \label{09}
\begin{array}{l}
\displaystyle{%
  \Delta\varphi=-8\pi|e|\big(|\psi_\alpha|^2-|\psi_e|^2\big).  %
}%
\end{array}
\end{equation}
The chemical potentials entering into (7),\,(8) can be expressed in
terms of the equilibrium density of the number of nuclei and
electron pairs $n_0=|\psi_{\alpha 0}|^2=|\psi_{e0}|^2$:
\begin{equation} \label{10}
\begin{array}{l}
\displaystyle{%
  \mu_\alpha=(g_\alpha+g_{\alpha e})n_0, \quad \mu_e=(g_e+g_{\alpha e})n_0.  %
}%
\end{array}
\end{equation}
The flux densities of the number of nuclei and electron pairs are
given by the formulas
\begin{equation} \label{11}
\begin{array}{l}
\displaystyle{%
  {\bf j}_\alpha=\frac{i\hbar}{2M}\big(\psi_\alpha\nabla\psi_\alpha^*-\psi_\alpha^*\nabla\psi_\alpha\big), \quad %
  {\bf j}_e=\frac{i\hbar}{2m}\big(\psi_e\nabla\psi_e^*-\psi_e^*\nabla\psi_e\big),  %
}%
\end{array}
\end{equation}
and the current densities of positive and negative charges: %
${\bf j}_{\alpha ch}=2|e|{\bf j}_\alpha$, ${\bf j}_{e ch}=-2|e|{\bf j}_e$. %
Thus, the equations (7)\,--\,(10) describe the dynamics of the
coherent system of nuclei and electron pairs and the electric
potential in such a system.\vspace{-2mm}

\section{Small oscillations of the coherent system of nuclei  and electron pairs}\vspace{-3mm}
Let us consider small oscillations in the spatially homogeneous
coherent system of nuclei and electron pairs in the absence of a
stationary flux, writing down complex functions in the form
\begin{equation} \label{12}
\begin{array}{l}
\displaystyle{%
  \psi_\alpha=\sqrt{n_0}+\delta\psi_\alpha, \quad \psi_e=\sqrt{n_0}+\delta\psi_e.  %
}%
\end{array}
\end{equation}
In the following, instead of the complex quantities
$\delta\psi_\alpha, \delta\psi_e$, it will be more convenient to use
the real functions
\begin{equation} \label{13}
\begin{array}{l}
\displaystyle{%
  \delta\Psi_\alpha=\delta\psi_\alpha+\delta\psi_\alpha^*, \quad \delta\Phi_\alpha=i\big(\delta\psi_\alpha-\delta\psi_\alpha^*\big),  %
}\vspace{2mm}\\ %
\displaystyle{%
  \delta\Psi_e=\delta\psi_e+\delta\psi_e^*, \quad\,\, \delta\Phi_e=i\big(\delta\psi_e-\delta\psi_e^*\big).  %
}%
\end{array}
\end{equation}
The fluctuations of the density of the number of nuclei $\delta
n_\alpha$, the density of the number of electron pairs $\delta n_e$,
the density of mass $\delta\rho_m$ and charge $\delta\rho_{ch}$, as
well as the fluctuations of the flux densities in terms of the
quantities (13) are given by the expressions\vspace{-1mm}
\begin{equation} \label{14}
\begin{array}{ccc}
\displaystyle{%
  \delta n_\alpha=\sqrt{n_0}\,\delta\Psi_\alpha, \quad \delta n_e=\sqrt{n_0}\,\delta\Psi_e, %
}\vspace{2mm}\\ %
\displaystyle{%
  \delta\rho_m=\sqrt{n_0}\big(M_\alpha\delta\Psi_\alpha+2m_e\delta\Psi_e\big), \quad \delta\rho_{ch}=2|e|\sqrt{n_0}\big(\delta\Psi_\alpha-\delta\Psi_e\big),    %
}\vspace{2mm}\\ %
\displaystyle{%
  \delta{\bf j}_\alpha=-\frac{\hbar\sqrt{n_0}}{2M}\nabla\delta\Phi_\alpha,\quad  \delta{\bf j}_e=-\frac{\hbar\sqrt{n_0}}{2m}\nabla\delta\Phi_e,\quad \delta{\bf j}_{ch}=2|e|\big(\delta{\bf j}_\alpha-\delta{\bf j}_e\big).   %
}%
\end{array}
\end{equation}

The linearized system of equations (7)\,--\,(9) for the real
variables (13) has the form\vspace{-1mm}
\begin{equation} \label{15}
\begin{array}{l}
\displaystyle{%
  \hbar\,\frac{\partial \delta\Phi_\alpha}{\partial t}=-\frac{\hbar^2}{2M}\Delta\delta\Phi_\alpha+2g_\alpha n_0\delta\Psi_\alpha+2g_{\alpha e}n_0\delta\Psi_e+2|e|\sqrt{n_0}\,\varphi, %
}
\end{array}
\end{equation}
\vspace{-7mm}
\begin{equation} \label{16}
\begin{array}{l}
\displaystyle{%
  \hbar\,\frac{\partial \delta\Psi_\alpha}{\partial t}=\frac{\hbar^2}{2M}\Delta\delta\Phi_\alpha, %
}
\end{array}
\end{equation}
\vspace{-7mm}
\begin{equation} \label{17}
\begin{array}{l}
\displaystyle{%
  \hbar\,\frac{\partial \delta\Phi_e}{\partial t}=-\frac{\hbar^2}{2m}\Delta\delta\Phi_e+2g_e n_0\delta\Psi_e+2g_{\alpha e}n_0\delta\Psi_\alpha-2|e|\sqrt{n_0}\,\varphi, %
}
\end{array}
\end{equation}
\vspace{-7mm}
\begin{equation} \label{18}
\begin{array}{l}
\displaystyle{%
  \hbar\,\frac{\partial \delta\Psi_e}{\partial t}=\frac{\hbar^2}{2m}\Delta\delta\Phi_e, %
}
\end{array}
\end{equation}
\vspace{-6mm}
\begin{equation} \label{19}
\begin{array}{l}
\displaystyle{%
  \Delta\varphi=-8\pi|e|\sqrt{n_0}\big(\delta\Psi_\alpha-\delta\Psi_e\big). %
}
\end{array}
\end{equation}
This system of five equations is equivalent to the system of two
equations for the functions $\delta\Psi_\alpha$ and $\delta\Psi_e$: %
\begin{equation} \label{20}
\begin{array}{l}
\displaystyle{%
  \frac{\partial^2\delta\Psi_\alpha}{\partial t^2}=-\frac{\hbar^2}{4M^2}\Delta^2\delta\Psi_\alpha+\frac{g_\alpha n_0}{M}\Delta\delta\Psi_\alpha+\frac{g_{\alpha e} n_0}{M}\Delta\delta\Psi_e-\omega_\alpha^2\delta\Psi_\alpha+\omega_\alpha^2\delta\Psi_e,  %
}%
\end{array}
\end{equation}
\vspace{-4mm}%
\begin{equation} \label{21}
\begin{array}{l}
\displaystyle{%
  \frac{\partial^2\delta\Psi_e}{\partial t^2}=-\frac{\hbar^2}{4m^2}\Delta^2\delta\Psi_e+\frac{g_e n_0}{m}\Delta\delta\Psi_e+\frac{g_{\alpha e} n_0}{m}\Delta\delta\Psi_\alpha+\omega_e^2\delta\Psi_\alpha-\omega_e^2\delta\Psi_e.  %
}%
\end{array}
\end{equation}
Here the plasma frequencies for nuclei $\omega_\alpha$ and electron
pairs $\omega_e$ are determined by the relations
\begin{equation} \label{22}
\begin{array}{l}
\displaystyle{%
   \omega_\alpha^2=\frac{8\pi e^2n_0}{M}, \quad \omega_e^2=\frac{8\pi e^2n_0}{m}.  %
}%
\end{array}
\end{equation}
\begin{figure}[b]
\vspace{0mm} \hspace{-0mm}
\includegraphics[width = 8.0cm]{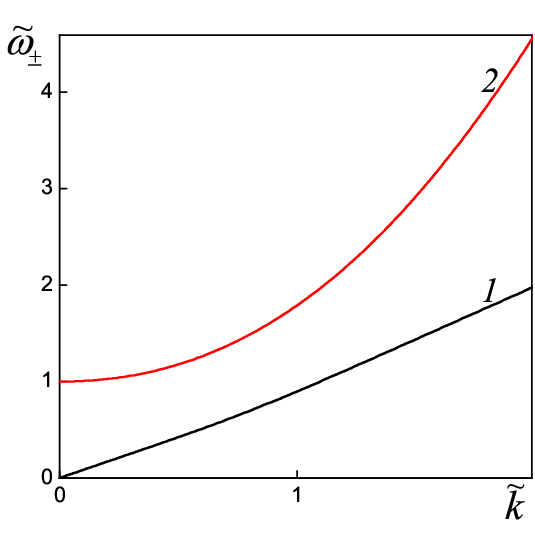} 
\vspace{-4mm}
\caption{\label{fig01} %
The sound ({\it 1}) and optical ({\it 2}) branches of excitations in
the system
with two oppositely charged coherent Bose-Einstein condensates. %
Here $\tilde{\omega}\equiv\omega/\omega_0$, $\tilde{k}\equiv k/k_0$, %
$\omega_0^2\equiv\omega_\alpha^2+\omega_e^2$, $k_0^2\equiv\frac{2Mm}{\sqrt{M^2+m^2}}\frac{\omega_0}{\hbar}$, %
$\gamma\equiv m/M=0.1.$
}%
\end{figure}\hspace{-1.5mm} %
Assuming the dependencies of the quantities
$\delta\Psi_\alpha$ and
$\delta\Psi_e$ on coordinates and time in the form $\exp i(\omega t-{\bf k}{\bf r})$, %
we find from (20) and (21) the dispersion equation
\begin{equation} \label{23}
\begin{array}{l}
\displaystyle{%
   \omega^4-2B\omega^2+C=0,  %
}%
\end{array}
\end{equation}
where
\begin{equation} \label{24}
\begin{array}{l}
\displaystyle{%
  2B=\omega_\alpha^2+\omega_e^2 +\left(\frac{g_\alpha}{M}+\frac{g_e}{m}\right)n_0k^2+\left(\frac{1}{4M^2}+\frac{1}{4m^2}\right)\hbar^2k^4, %
}\vspace{3mm}\\ %
\displaystyle{%
  C=\left[\omega_\alpha^2\frac{(g_e+g_{\alpha e})}{m}+\omega_e^2\frac{(g_\alpha+g_{\alpha e})}{M}\right]\!n_0k^2 +\! %
  \left[\hbar^2\!\left(\frac{\omega_\alpha^2}{4m^2}+\frac{\omega_e^2}{4M^2}\right)+\frac{\big(g_\alpha g_e-g_{\alpha e}^2\big)}{mM}n_0^2\right]\!k^4 +%
}\vspace{1mm}\\ %
\displaystyle{%
\hspace{5mm}  +\left(\frac{g_\alpha}{4Mm^2}+\frac{g_e}{4mM^2}\right)\!n_0\hbar^2k^6 +\frac{\hbar^4k^8}{16M^2m^2}\,. %
}%
\end{array}
\end{equation}
Thus, there are two branches of excitations
\begin{equation} \label{25}
\begin{array}{l}
\displaystyle{%
   \omega_\pm^2=B\pm\sqrt{B^2-C},  %
}%
\end{array}
\end{equation}
which are shown in Fig.\,1. In the short-wavelength limit, these
branches transform into the dispersion laws of free nuclei and
electron pairs
\begin{equation} \label{26}
\begin{array}{l}
\displaystyle{%
   \omega_+=\frac{\hbar k^2}{2m}, \quad \omega_-=\frac{\hbar k^2}{2M}.  %
}%
\end{array}
\end{equation}
These limiting relations seem reasonable, but it is physically
correct to consider the dispersion relations in the limit of long
waves. In this case
\begin{equation} \label{27}
\begin{array}{l}
\displaystyle{%
  \omega_\pm^2=\frac{1}{2}\Big[\omega_\alpha^2+\omega_e^2 +\left(\frac{g_\alpha}{M}+\frac{g_e}{m}\right)\!n_0k^2\Big]\pm %
  \frac{1}{2}\left\{\Big[\omega_\alpha^2+\omega_e^2 +\left(\frac{g_\alpha}{M}+\frac{g_e}{m}\right)\!n_0k^2\Big]^2-\frac{4\big(g_\alpha g_e-g_{\alpha e}^2\big)n_0^2k^4}{mM}\right\}^{\!1/2}. %
}
\end{array}
\end{equation}
In the system of two neutral condensates, at $e=0$, we have two
sound branches $\omega_\pm^2=c_{0\pm}^2k^2$, where
\begin{equation} \label{28}
\begin{array}{l}
\displaystyle{%
   c_{0\pm}^2=\frac{n_0}{2}\!\left[\left(\frac{g_\alpha}{M}+\frac{g_e}{m}\right)\pm\sqrt{\left(\frac{g_\alpha}{M}+\frac{g_e}{m}\right)^{\!2}-\frac{4\big(g_\alpha g_e-g_{\alpha e}^2\big)}{mM}}\,\right]\!.  %
}%
\end{array}
\end{equation}

In the case of charged condensates we are interested in, there is a
single sound branch $\omega_-=ck$, where the velocity is determined
by the formula
\begin{equation} \label{29}
\begin{array}{l}
\displaystyle{%
  c^2=\frac{n_0(g_\alpha+g_e+2g_{\alpha e})}{(m+M)}. %
}%
\end{array}
\end{equation}
For stability of the system, the interaction constants must satisfy
the condition $g_\alpha+g_e+2g_{\alpha e}>0$. The second branch is
optical $\omega_+^2=\omega_\alpha^2+\omega_e^2+\alpha k^2$, where
\begin{equation} \label{30}
\begin{array}{l}
\displaystyle{%
   \alpha=\frac{n_0}{\big(\omega_\alpha^2+\omega_e^2\big)}\!\left[\frac{g_\alpha}{M}\omega_\alpha^2+\frac{g_e}{m}\omega_e^2-g_{\alpha e}\!\left(\frac{\omega_\alpha^2}{m}+\frac{\omega_e^2}{M}\right)\right]=  %
   \frac{n_0}{(M+m)}\left(g_\alpha\frac{m}{M}+g_e\frac{M}{m}-2g_{\alpha e}\right)\!. %
}%
\end{array}
\end{equation}
The gap in the spectrum $\omega_0$ is determined by the relation
\begin{equation} \label{31}
\begin{array}{l}
\displaystyle{%
  \omega_0^2=\omega_\alpha^2+\omega_e^2=\frac{8\pi n_0e^2}{M_*}, %
}%
\end{array}
\end{equation}
where $M_*=mM\big/(m+M)$ is the reduced mass. When deriving formulas
(29),\,(30) from the more general formula (27), it was assumed a
fulfillment of the condition
\begin{equation} \label{32}
\begin{array}{l}
\displaystyle{%
  k^2<\frac{(m+M)^2}{mM}\frac{e^2}{|g_i|}, \quad i=(\alpha,e,\alpha e). %
}%
\end{array}
\end{equation}

The proposed model can pretend to provide quantitative estimates
only in the case of low-density systems, whereas liquid helium is
not such one. Nevertheless, it is of interest to estimate the value
of the reduced mass, assuming that the formula (31) remains valid in
this case and the frequency $f_0=\omega_0/2\pi$ coincides with the
resonant frequency of 180 GHz observed in experiments [3--7]. At the
density $n_0=10^{22}\,\rm{cm}^{-3}$, it turns out that the reduced
mass is four orders of magnitude greater than the mass of a helium
atom: $M_*\propto 10^4M_{\rm{He}}$. Note that earlier it was drawn
attention to the possibility of the existence of a gap in the energy
spectrum of superfluid Bose systems due to pair correlations in
works [13,17,18,21]. The calculation of absorption of a microwave
radiation at the resonant frequency can be performed in a similar
way as in [22].

The fluctuations of the number of pairs $\delta n_e^{(-)}$ and
nuclei $\delta n_\alpha^{(-)}$ in the sound wave are linked by the relation %
\begin{equation} \label{33}
\begin{array}{l}
\displaystyle{%
   \delta n_e^{(-)}=\left[1-\frac{k^2}{8\pi e^2}\frac{\big(g_eM-g_\alpha m+g_{\alpha e}(M-m)\big)}{(M+m)}\right]\!\delta n_\alpha^{(-)}.  %
}%
\end{array}
\end{equation}
According to (14), the fluctuations of the mass density and charge
density in the sound wave are given by the formulas
\begin{equation} \label{34}
\begin{array}{l}
\displaystyle{%
   \delta \rho_m^{(-)}=\left[M_\alpha+2m_e-m_e\frac{k^2}{4\pi e^2}\frac{\big(g_eM-g_\alpha m+g_{\alpha e}(M-m)\big)}{(M+m)}\right]\!\delta n_\alpha^{(-)},  %
}%
\end{array}
\end{equation}
\vspace{-3mm}
\begin{equation} \label{35}
\begin{array}{l}
\displaystyle{%
   \delta \rho_{ch}^{(-)}=\frac{k^2}{4\pi |e|}\frac{\big(g_eM-g_\alpha m+g_{\alpha e}(M-m)\big)}{(M+m)}\,\delta n_\alpha^{(-)}.  %
}%
\end{array}
\end{equation}
From (34) and (35) there follows the relation between the
fluctuations of charge and mass densities
\begin{equation} \label{36}
\begin{array}{l}
\displaystyle{%
   \delta \rho_{ch}^{(-)}=\frac{k^2}{4\pi |e|}\frac{\big(g_eM-g_\alpha m+g_{\alpha e}(M-m)\big)}{M_{\rm{He}}(M+m)}\,\delta\rho_m^{(-)},  %
}%
\end{array}
\end{equation}
where $M_{\rm{He}}\equiv M_\alpha+2m_e$ is the mass of a helium atom
$^4$He. As we can see, at $k\rightarrow 0$ also $\delta
\rho_{ch}^{(-)}\rightarrow 0$, so that with an increase in the
length of the sound wave the charge fluctuation in it decreases in
comparison to the density fluctuation. It should be noted, however,
that the wavelength cannot exceed the characteristic size of the
system, so that always $k\geq 1/L$. This is essential, as will be
seen below, when considering oscillations in a resonator. Let us
also give the relation between the current density fluctuation and
the mass density fluctuation:
\begin{equation} \label{37}
\begin{array}{l}
\displaystyle{%
   \delta{\bf j}_{ch}^{(-)}=\frac{\hbar\,|e|}{M_{\rm{He}}}\!\left(\frac{1}{m}-\frac{1}{M}\right)\!\nabla\delta \rho_m^{(-)}. %
}%
\end{array}
\end{equation}

For the optical branch, the fluctuations of the number of pairs
$\delta n_e^{(+)}$ and nuclei $\delta n_\alpha^{(+)}$ are linked by
the relation
\begin{equation} \label{38}
\begin{array}{l}
\displaystyle{%
   \delta n_e^{(+)}=-\frac{M}{m}\!\left[1+\frac{k^2}{8\pi e^2}\frac{\big(g_eM-g_\alpha m+g_{\alpha e}(M-m)\big)}{(M+m)}\right]\!\delta n_\alpha^{(+)}.  %
}%
\end{array}
\end{equation}
According to (14), the oscillations of the mass and charge densities
in the optical branch are determined by the formulas
\begin{equation} \label{39}
\begin{array}{l}
\displaystyle{%
   \delta \rho_m^{(+)}=\left[M_\alpha-2\frac{m_e}{m}M-\frac{m_eM}{m}\frac{k^2}{4\pi e^2}\frac{\big(g_eM-g_\alpha m+g_{\alpha e}(M-m)\big)}{(M+m)}\right]\!\delta n_\alpha^{(+)},  %
}%
\end{array}
\end{equation}
\vspace{-3mm}
\begin{equation} \label{40}
\begin{array}{l}
\displaystyle{%
   \delta \rho_{ch}^{(+)}=2\frac{|e|}{m}\!\left[(M+m)+\frac{k^2}{8\pi |e|}\frac{M\big(g_eM-g_\alpha m+g_{\alpha e}(M-m)\big)}{(M+m)}\right]\!\delta n_\alpha^{(+)}.  %
}%
\end{array}
\end{equation}
From (39) and (40) it follows that in this case the relation between
the fluctuations of charge and mass densities is given by the
formula
\begin{equation} \label{41}
\begin{array}{l}
\displaystyle{%
   \delta \rho_{ch}^{(+)}=\frac{2\,|e|(M+m)}{(mM_\alpha-2m_eM)}\!\left[1+\frac{k^2}{8\pi |e|}\frac{M\big(g_eM-g_\alpha m+g_{\alpha e}(M-m)\big)}{(M+m)^2}\right]\!\delta\rho_m^{(+)}.  %
}%
\end{array}
\end{equation}
In the limit $k\rightarrow 0$, the ratio of the amplitudes of
oscillations of charge and mass densities remains constant. The
current density fluctuation and the mass density fluctuation are
linked by the relation
\begin{equation} \label{42}
\begin{array}{l}
\displaystyle{%
   \delta{\bf j}_{ch}^{(+)}=-\hbar\,|e|\frac{\left(\frac{M}{m}+\frac{m}{M}\right)}{(mM_\alpha-2m_eM)}\,\nabla\delta\rho_m^{(+)}. %
}%
\end{array}
\end{equation}

Although this work does not consider the states with stationary
flows, we note that since electron pairs compensate for the charge of nuclei,
stationary flux densities of the number of nuclei and pairs should
be the same and the stationary electric current density should be
zero in this case.

\section{Low frequency oscillations in a capacitor}
In the experiment [1], there were studied the standing waves of the
second sound and the potential oscillations in a resonator filled
with superfluid helium. Let us consider, within the framework of the
proposed model, the oscillations in a capacitor the plates of which
are perpendicular to the $x$-axis and located at the points $x=\pm
L/2$. Taking into account that the flows of particles in the
direction perpendicular to the plates should vanish on the plates
themselves, we find that the fluctuations of the densities of nuclei
and pairs are given by the formulas
\begin{equation} \label{43}
\begin{array}{l}
\displaystyle{%
   \delta n_\alpha=\overline{n}_\alpha\sin\omega_n t\sin\!k_nx, \quad %
   \delta n_e=\overline{n}_e\sin\omega_n t\sin\!k_nx,  %
}%
\end{array}
\end{equation}
where $\omega_n=ck_n$, $k_n=\pi(2n+1)/L$, and the velocity $c$ is
determined by the formula (29). Assuming that the potential on the
right plate at $x=L/2$ is equal to zero and the surface charges on
the capacitor plates are absent, so that the normal component of the
electric field vanishes on the plates, we find from the equation
(19) the potential distribution
\begin{equation} \label{44}
\begin{array}{l}
\displaystyle{%
   \delta\varphi(x,t)=\frac{\varphi_m}{2}\big[(-1)^n-\sin\!k_nx\big]\sin\omega_n t, %
}%
\end{array}
\end{equation}
where $\displaystyle{ \varphi_m\equiv -\frac{2\big(g_eM-g_\alpha m+g_{\alpha e}(M-m)\big)}{|e|(M+m)}\,\overline{n}_\alpha}$. %
The potential distribution for the cases when one $n=0$ and three
$n=1$ half-waves fit along the resonator length is shown in Fig.\,2.
\begin{figure}[t!]
\vspace{-12mm} \hspace{-0mm}
\includegraphics[width = 8.0cm]{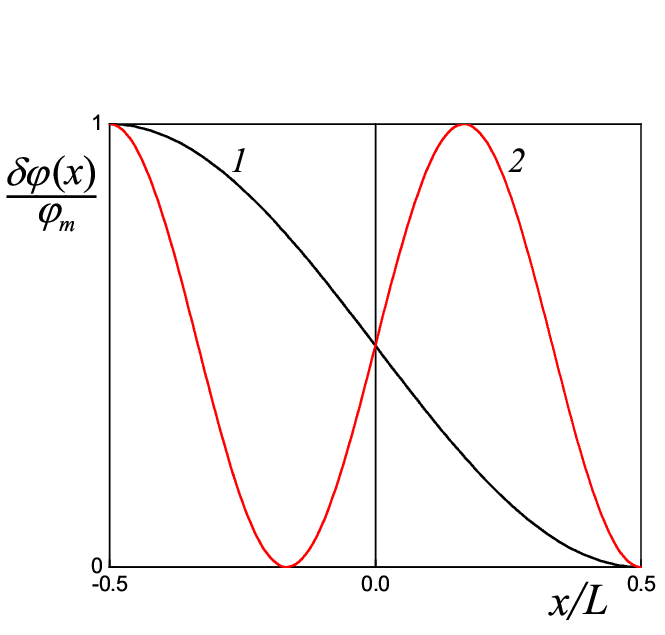} 
\vspace{-4mm}
\caption{\label{fig02} %
The potential distribution in the capacitor at the moment $\displaystyle{t=\frac{L}{2c}}$: %
({\it 1}) one $n=0$, ({\it 2}) three $n=1$ %
half-waves fit along the resonator length.
}%
\end{figure} 
These distributions coincide with those obtained experimentally in [20].\vspace{-2mm} %

\section{Discussion}\vspace{-2mm} %
In the semi-phenomenological model of helium superfluidity proposed
in this article, atoms are considered as complexes consisting of a
nucleus and a bound pair of electrons in the singlet state. Since
this model contains two systems of oppositely charged particles
obeying the Bose statistics, there are also two coherent
Bose-Einstein condensates. In dynamic processes the local
fluctuations of the number of electron pairs and nuclei lead to the
fluctuations of the densities of electric charge, current and
potential. The model is formulated for zero temperature and under
the assumption that the system is rarefied, and therefore cannot
pretend to give a quantitative description of the effects observed
in superfluid helium, like indeed any other model if it does not
contain a set of a sufficient number of adjustable parameters.
Nevertheless, it allows to qualitatively understand the cause of the
observed electrical phenomena, which consists in the perturbation of
the coherent system of nuclei and electron pairs. In liquid helium
the coherence emerges below the lambda-transition temperature. A
consequence of the emergence of the coherent Bose-Einstein
condensate is the appearance of a new characteristic of the system
-- the superfluid density. Thus, in liquid superfluid helium the
coherent subsystem of atoms forms the superfluid density, which
depends on temperature $T$, pressure $p$ and the difference of the
superfluid and normal velocities ${\bf w}={\bf v}_n-{\bf v}_s$, so
that $\rho_s=\rho_s(p,T,w^2)$. As we can see, the oscillations of
the superfluid density, and consequently of the coherent subsystem,
in helium can arise under the influence of the fluctuations of
temperature, pressure and the velocity difference. The intensity of
such oscillations of the coherent subsystem can be characterized by
the dimensionless parameters
\begin{equation} \label{45}
\begin{array}{l}
\displaystyle{%
  A_p=\frac{p}{\rho_s}\left(\frac{\partial\rho_s}{\partial p}\right)_{\!T,w}, \quad  %
  A_T=\frac{T}{\rho_s}\left(\frac{\partial\rho_s}{\partial T}\right)_{\!p,w}, \quad  %
  A_w=\frac{u_2^2}{\rho_s}\left(\frac{\partial\rho_s}{\partial w^2}\right)_{\!T,p},   %
}%
\end{array}
\end{equation}
where $u_2$ is the velocity of the second sound.     Let us estimate
the magnitude of these coefficients. Using the data given in the
appendix of the book [23], we find that at $T=1.4\,{\rm K}$ the
coefficient $A_T\approx -0.54$. With an increase in temperature,
this coefficient increases in absolute magnitude, reaching at
$T=2\,{\rm K}$ the value $A_T\approx -7.3$. In this case, of course,
$\rho_s$ decreases.

Let us also estimate the magnitude of the coefficient $A_w$ that
determines the influence of the oscillations of the superfluid flow
on the perturbation of the coherent system of helium. The derivative
of the superfluid density is expressed in terms of the derivatives
of the total $\rho$ and normal $\rho_n$ densities: %
$\partial\rho_s/\partial w^2=\partial\rho/\partial w^2-\partial\rho_n/\partial w^2$. %
The first term can be estimated using the thermodynamic relation [24] %
\begin{equation} \label{46}
\begin{array}{l}
\displaystyle{%
  \frac{\partial\rho}{\partial w^2}=\frac{\rho^2}{2}\frac{\partial}{\partial p}\!\left(\frac{\rho_n}{\rho}\right)\!.   %
}%
\end{array}
\end{equation}
The dependence of the normal density on $w^2$ is found from formulas
given in $\S\,3$ of [24]. As a result, we get
\begin{equation} \nonumber
\begin{array}{l}
\displaystyle{%
  \frac{\partial\rho_s}{\partial w^2}\approx -0.85\cdot\!10^{-9}\,\rm{g\!\cdot\! s^2}/\rm{cm^5}.   %
}%
\end{array}
\end{equation}
This estimate is consistent with the experimental estimate given in
the appendix of the book [23]\,
$\rho^{-1}\!\left|\partial\rho_n\big/\partial w^2\right|< 6\cdot\!10^{-8}\,\rm{s^2}/\rm{cm^2}$. %
Taking into account the value of the second-sound velocity $u_2=2\cdot\!10^{3}\,\rm{cm}/\rm{s}$, %
we get $A_w\approx -2.5\cdot\!10^{-2}$. And if we take the maximum possible value %
$\rho^{-1}\!\left|\partial\rho_n\big/\partial w^2\right|= 6\cdot\!10^{-8}\,\rm{s^2}/\rm{cm^2}$ %
according to the experimental data, then we get $A_w\approx -0.25$.
As seen, the electrical effects caused by the fluctuations of ${\bf
w}={\bf v}_n-{\bf v}_s$ are close to those generated by the
temperature fluctuations. The electrical effects caused by the
fluctuations of the velocity difference were observed in the
experiment with a torsion oscillator [2].

An estimate of the coefficient that determines the effect of
pressure on the coherent subsystem at saturated vapor pressure gives
$A_p\approx 10^{-5}$. Pressure has the least effect on the coherent
subsystem, so that the electrical effects should be much less
pronounced in experiments with the first sound. However, as the
pressure increases the coefficient $A_p$ also increases. So, at
pressure of 5\,\,atm  it has an order of magnitude $A_p\approx
10^{-2}$. Note that in [25] the observation of the electric effect
in the first sound wave was reported, although this effect was not
observed in other works.\vspace{-1mm}

\section{Conclusion}\vspace{-2mm} %
The article proposes a qualitative interpretation of the electrical
effects observed in superfluid helium [1--12,20,25], based on the
analysis of the model which assumes the existence of two oppositely
charged coherent Bose-Einstein condensates -- those of atomic nuclei
and singlet electron pairs. In this approach the electron pairs are
considered as delocalized, so that in nonstationary processes
there exists a probability of a pair transition from atom to
atom and, therefore, the possibility of the local breaking of
electroneutrality, which thus leads to the appearance of the
internal electric field. It is shown that there are two branches of
elementary excitations -- sound and optical.

The observed electrical activity in superfluid helium is explained
by the disturbance due to external factors of its coherent system
manifesting itself in the existence of the superfluid density.
Estimates show that the strongest effect on the coherent system is
exerted by the fluctuations of temperature, then follow the
fluctuations of superfluid flow, and the weakest effect is due to
the fluctuations of pressure. The frequency at which the resonant
absorption of a microwave radiation is observed [3--7] is
interpreted as a gap in the optical branch of the spectrum. The
oscillations in a resonator are considered and it is shown that the
distribution of the electric potential in the standing wave is
consistent with experiment [20].


\begin{thebibliography}{99}
\bibitem{Rybalko1}
A.S.\,Rybalko, Low Temp. Phys. \textbf{30}, 994 (2004).
$\mbox{doi:10.1063/1.1820042}$
\bibitem{Rybalko2}
A.S.\,Rybalko, S.P.\,Rubets, Low Temp. Phys. \textbf{31}, 623 (2005). %
$\mbox{doi:10.1063/1.2001649}$
\bibitem{Rybalko3}
A.S.\,Rybalko, S.P.\,Rubets, E.Ya.\,Rudavskii, V.A.\,Tikhiy,
S.I.\,Tarapov, R.V.\,Golovashchenko, V.N.\,Derkach, Phys. Rev. B \textbf{76}, 140503(R) (2007). %
$\mbox{doi:10.1103/PhysRevB.76.140503}$
\bibitem{Rybalko4}
A.S.\,Rybalko, S.P.\,Rubets, E.Ya.\,Rudavskii, V.A.\,Tikhiy,
R.V.\,Golovashchenko, V.N.\,Derkach, S.I.\,Tarapov, Low Temp. Phys. \textbf{34}, 254 (2008). %
$\mbox{doi:10.1063/1.2911649}$
\bibitem{Rybalko5}
A.S.\,Rybalko, S.P.\,Rubets, E.Ya.\,Rudavskii, V.A.\,Tikhiy,
S.I.\,Tarapov, R.V.\,Golovashchenko, V.N.\,Derkach, Low Temp. Phys. \textbf{34}, 497 (2008). %
$\mbox{doi:10.1063/1.2957000}$
\bibitem{Rybalko6}
A.S.\,Rybalko, S.P.\,Rubets, E.Ya.\,Rudavskii, V.A.\,Tikhiy,
Yu.M.\,Poluektov, R.\,Golovashchenko, V.N.\,Derkach,
S.I.\,Tarapov, O.V.\,Usatenko, Low Temp. Phys. \textbf{35}, 837 (2009). %
$\mbox{doi:10.1063/1.3266909}$
\bibitem{Rybalko7}
A.S.\,Rybalko, S.P.\,Rubets, E.Ya.\,Rudavskii, V.A.\,Tikhiy,
Yu.M.\,Poluektov, R.\,Golovashchenko, V.N.\,Derkach,
S.I.\,Tarapov, O.V.\,Usatenko, J. Low Temp. Phys. \textbf{158}, 244 (2010). %
$\mbox{doi:10.1007/s10909-009-0025-6}$
\bibitem{Chagovets1}
T.V.\,Chagovets, Low Temp. Phys. \textbf{42}, 176 (2016). %
$\mbox{doi:10.1063/1.4942758}$
\bibitem{Chagovets2}
T.V.\,Chagovets, Physics B \textbf{488}, 62 (2016). %
$\mbox{doi:10.1016/j.physb.2016.02.008}$
\bibitem{Rybalko8}
A.S.\,Rybalko, T.V.\,Chagovets, A.M.\,Korolev, Low Temp. Phys. \textbf{43}, 738 (2017). %
$\mbox{doi:10.1063/1.4985984}$
\bibitem{Yayama}
H.\,Yayama, Y.\,Nishimura, H.\,Ushiyama, H.\,Kawai, J.-P. van
Woensel, A. Hafez, Low Temp. Phys. \textbf{44}, 1090 (2018). %
$\mbox{doi:10.1063/1.5055857}$
\bibitem{Natsik}
V.D.\,Natsik, A.S.\,Rybalko, Low Temp. Phys. \textbf{46}, 28 (2020). %
$\mbox{doi:10.1063/10.0000361}$
\bibitem{Poluektov1}
Yu.M.\,Poluektov, Low Temp. Phys. \textbf{37}, 986 (2011). %
$\mbox{doi:10.1063/1.3674269}$
\bibitem{Shevchenko}
S.I.\,Shevchenko, A.S.\,Rukin, Low Temp. Phys. \textbf{38}, 905 (2012). %
$\mbox{doi:10.1063/1.4758763}$
\bibitem{Gross}
E.P.\,Gross, Nuovo Cimento \textbf{20}, 454 (1961).
$\mbox{doi:10.1007/BF02731494}$
\bibitem{Pitaevskii}
L.P.\,Pitaevskii, Sov. Phys. JETP \textbf{13}, 451 (1961). %
\bibitem{PPP1}
A.S.\,Peletminskii, S.V.\,Peletminskii, Yu.M.\,Poluektov, Condensed
Matter Physics \textbf{16}, 13603 (2013) [arXiv:1303.5539]. %
$\mbox{doi:10.5488/CMP.16.13603}$
\bibitem{Poluektov2}
Yu.M.\,Poluektov, Low Temp. Phys. \textbf{44}, 1040 (2018). %
$\mbox{doi:10.1063/1.5055845}$
\bibitem{Pashitskii}
E.A.\,Pashitskii, S.M.\,Ryabchenko, Low Temp. Phys. \textbf{33}, 8 (2007). %
$\mbox{doi:10.1063/1.2409628}$
\bibitem{Rybalko9}
A.S.\,Rybalko, V.A.\,Tikhiy, A.S.\,Neoneta, K.R.\,Zhekov,
arXiv:1108.1272v1\,[cond-mat.other] (2011).
\bibitem{PPP2}
A.S.\,Peletminskii, S.V.\,Peletminskii, Yu.M.\,Poluektov, Low Temp. Phys. \textbf{40}, 500 (2014). %
$\mbox{doi:10.1063/1.4883893}$
\bibitem{Poluektov3}
Yu.M.\,Poluektov, Low Temp. Phys. \textbf{40}, 389 (2014). %
$\mbox{doi:10.1063/1.4881395}$
\bibitem{Patterman}
S.J.\,Patterman, Hydrodynamics of Superfluid Liquid, Mir, Moscow, 520\,p. (1978). %
\bibitem{Khalatnikov}
I.M.\,Khalatnikov, Theory of superfluidity, Nauka, Moscow, 320\,p. (1971). %
\bibitem{Chagovets3}
T.V.\,Chagovets, J. Low Temp. Phys. \textbf{187}, 383 (2017). %
$\mbox{doi:10.1007/s10909-017-1746-6}$
\end{thebibliography}
\end{document}